\documentclass{article}
\newcommand{\bb}{\mbox{\boldmath{$b$}}}

\newcommand{\by}{\mbox{\boldmath{$y$}}}

\begin{document}
\noindent
{\footnotesize{\bf
Randomness and Computation \\
Joint Workshop
``New Horizons in Computing'' and 
``Statistical Mechanical Approach to Probabilistic
Information Processing'' (18-21 July, 2005, Sendai, Japan)}}

\vspace*{1.0cm}
\begin{center}
\LARGE \bf 
A CDMA multiuser detection algorithm based on survey propagation 
\end{center}
\vspace*{0.25cm}

\begin{center}
{\large Yoshiyuki Kabashima} 
\footnote{E-mail: kaba@dis.titech.ac.jp} 
\\
Department of Computational Intelligence and Systems Science, 
Tokyo Institute of Technology, 
Yokohama 226-8502, Japan 
\end{center}

Code division multiple access (CDMA) is a core technology 
of today's wireless communication system that employs data transmission 
between multiple terminals and a single base station. 
In the general scenario of a CDMA system, 
the binary signals of multiple users are modulated 
by spreading codes assigned to each user,
and these modulated sequences are transmitted to a base station. 
The base station receives real-valued signals, which are 
a superposition of the modulated sequences and possible noise. 
A detector at the base station then extracts the original 
binary signals from the received signals using knowledge
of the spreading codes of the users. This task is termed 
{\em user detection}.

The Bayesian framework indicates that 
the optimal detection performance can be 
achieved by multiuser detection, which simultaneously
infers the signals of multiple users. 
Although performing such optimal detection 
exactly is computationally difficult, 
this difficulty can be approximately 
resolved by employing belief propagation (BP),
and the macroscopic behavior of the 
BP-based algorithm is well captured 
by the natural iteration of the saddle point 
equations of replica symmetric (RS)
analysis~\cite{CDMABP}. Unfortunately, the performance of the 
BP-based algorithm deteriorates as the discrepancy 
between the true and assumed channel models becomes larger. 
In particular, this algorithm exhibits {\em dynamic}
instability when the intensity of the channel noise 
is set below a certain critical value, which is in accordance
with the de Almeida-Thouless (AT) instability in 
{\em equilibrium} analysis~\cite{CDMABP}. 
This correspondence of two types of instability has motivated the development of a multiuser detection algorithm based on survey propagation 
(SP), which is a recently 
developed algorithm that is inspired by the one-step replica symmetry 
breaking (1RSB) solution of equilibrium analysis~\cite{SP}. 
This is the purpose of the present 
study\footnote{
A similar but different approach 
to an improved algorithm can be found in \cite{Saad}.}.

We focus on a CDMA system that uses binary phase shift keying 
(BPSK) symbols and $K$ random binary spreading codes of 
the spreading factor $N$ with unit energy over 
an additive white Gaussian noise (AWGN) channel. 
For simplicity, we assume that the signal powers
are completely controlled to unit energy, but the extension to 
the case of distributed power is straightforward. 
In addition, we assume that chip timing and symbol 
timing are perfectly synchronized among users. Under these 
assumptions~\cite{tanaka}, the received 
real-valued signal can be expressed as
\begin{eqnarray}
y_\mu=\frac{1}{\sqrt{N}}\sum_{k=1}^Ks_{\mu k} b_k^0 + \sigma_0 n_\mu,  
\label{eq:base_band}
\end{eqnarray}
where $\mu \in \{1,2,\ldots,N\}$ and $k \in \{
1,2,\ldots,K\}$ are indices of samples and users, 
respectively, $s_{\mu k} \in \{-1,1\}$ is the spreading code 
with unit energy independently generated from the identical 
unbiased distribution $P(s_{\mu k}=+1)=P(s_{\mu k}=-1)=1/2$, 
$b_k^0$ is the bit signal of user $k$, $n_\mu$ is 
a Gaussian white noise 
sample with zero mean and unit variance, and 
$\sigma_0$ is the standard deviation of AWGN. 

In the Bayesian framework, the multiuser detection problem 
can be formulated as a problem to infer $K$ bit user-signals
$\bb^0=(b_k^0)$ 
from the posterior distribution 
$P(\bb|\by)=\frac{\prod_{\mu=1}^N P(y_\mu|\bb)}
{\sum_{\bb} \prod_{\mu=1}^N P(y_\mu|\bb)}$, 
with knowledge of the spreading codes 
of the users $\{s_{\mu k}\}$, 
where
$P(y_\mu|\bb) = \frac{1}{\sqrt{2 \pi \sigma^2}}
\exp \left [ -\frac{1}{2 \sigma^2} 
\left (y_\mu - u_\mu \right )^2 \right ]$, 
$\bb=(b_k)=\{+1,-1\}^K$, 
$u_\mu \equiv \frac{1}{\sqrt{N}}
\sum_{k=1}^K s_{\mu k}b_k$
and $\sigma$ is the noise parameter assumed
at the base station. 

For such an inference problem, the replicated Bethe free energy 
formalism~\cite{Kaba_SP} expresses SP as 
\begin{eqnarray}
&&\hat{\pi}^{t+1}_{\mu \to k}(\hat{h})
\propto  \int \prod_{j  \ne  k}
dh_j \pi^t_{j \to \mu}(h_j) 
\left (\sum_{\bb} P(y_\mu|\bb) 
\prod_{j \ne k}
\left (\frac{e^{h_jb_j}}{2\cosh (h_j)} \right ) \right )^x
\delta 
\left (\hat{h}-\hat{h}(\{h_{j \ne k}
 \}) \right ), 
\label{SP1}\\
&&\pi^t_{k \to \mu}(h)
\propto 
(2 \cosh (h))^x 
\int \prod_{\nu \ne \mu}
\frac{d\hat{h}_\nu \hat{\pi}^t_{\nu \to k}(\hat{h}_\nu)}
{(2 \cosh (\hat{h_\nu}))^x }
\delta (h-\sum_{\nu \ne \mu}
\hat{h}_\nu ), 
\label{SP2}
\end{eqnarray}
where $t$ represents the number of updates
and auxiliary field distributions
$\pi_{k \to \mu}^t(h)$ and $\hat{\pi}_{\mu \to k}^t(\hat{h})$
are termed {\em surveys}, the stationary solutions 
of which characterize the replica symmetric stationary 
points of the replicated Bethe free energy. 
$\hat{h}(\{h_{j \ne  k} \})=\tanh^{-1}\left (
\frac{\sum_{\bb}b_k P(y_\mu|\bb) 
\prod_{j \ne k}
\left (\frac{e^{h_jb_j}}{2
\cosh (h_j)} \right )}
{\sum_{\bb} P(y_\mu|\bb) 
\prod_{j \ne k}
\left (\frac{e^{h_jb_j}}{2\cosh (h_j)} \right )} \right )
$ and $x$ corresponds to the replica symmetry breaking 
parameter of the 1RSB analysis.
At each update, the marginal distribution 
is approximately assessed as 
$P(b_k|\by)=\sum_{b_{j \ne k}}P(\bb|\by)\simeq 
\int dh \rho_k(h) \frac{e^{hb_k}}{2 \cosh (h_k)}, 
$
where 
$\rho_k(h)=\frac{(2 \cosh (h))^x \int 
\prod_{\mu =1}^N
\frac{d\hat{h}_\mu \hat{\pi}_{\mu \to k}(\hat{h}_\mu)}{
(2\cosh (\hat{h}_\mu))^x} 
\delta(h-\sum_{\mu =1 }^N
\hat{h}_\mu) }
{ \int 
\prod_{\mu =1}^N
\frac{d\hat{h}_\mu \hat{\pi}_{\mu \to k}(\hat{h}_\mu)}{
(2\cosh (\hat{h}_\mu))^x} 
(2 \cosh (\sum_{\mu =1 }^N
\hat{h}_\mu))^x}$. 

Unfortunately, executing this algorithm exactly is 
practically difficult because the computational cost for 
assessing $\hat{h}(\{h_{j \ne  k} \})$ grows 
exponentially with respect to the number of users $K$. 
However, this difficulty can be overcome using Gaussian 
approximation when $K$ and $N$ are large while keeping 
$\beta=K/N \sim O(1)$, which leads to a computationally 
tractable algorithm 
\begin{eqnarray}
a_\mu^{t+1}=\frac{
\sigma^2 
\left (y_\mu-\frac{1}{\sqrt{N}}
\sum_{k=1}^Ns_{\mu k} m_k^t \right )+{\Xi^t} a_\mu^t}
{\sigma^2+\Xi^t}, \label{horizontal}\\
h_k^{t}(z)=\frac{1}{\sqrt{N}}\sum_{\mu=1}^N s_{\mu k} a_\mu^{t}
+\sqrt{\Delta^t}z+\Gamma^t m_k^{t-1},
\label{vertical}
\end{eqnarray}
where 
$\Delta^{t+1}=\frac{\beta(Q_1^t-Q_0^t)}{
(\sigma^2+\beta(1-Q_1^t))(\sigma^2+\beta(1-Q_1^t+x(Q_1^t-Q_0^t)))
}\sigma^2$, 
$\Gamma^{t+1}=\frac{\sigma^2}{
\sigma^2+\beta (1-Q_1^t+x(Q_1^t-Q_0^t))}$, 
$\Xi^{t}=\beta (1-Q_1^t+x(Q_1^t-Q_0^t))$,
$Q_0^t=K^{-1}\sum_{k=1}^K (m_k^t)^2$, 
$Q_1^t=K^{-1}\sum_{k=1}^K M_k^t$, 
$m_k^t=\frac{\int Dz (2 \cosh(\sigma^{-2}h_k^t(z)))^x 
\tanh(\sigma^{-2}h_k^t(z))}
{\int Dz (2 \cosh(\sigma^{-2}h_k^t(z)))^x}$
and 
$M_k^t=\frac{\int Dz (2 \cosh(\sigma^{-2}h_k^t(z)))^x
\tanh^2(\sigma^{-2}h_k^t(z))}{
\int Dz (2 \cosh(\sigma^{-2}h_k^t(z)))^x}$. 
Here, we set $Dz=\frac{dz}{\sqrt{2\pi}} 
e^{-\frac{1}{2}z^2}$. 
This provides the maximizer of posterior marginal (MPM) estimate, 
which minimizes the bit-wise detection error probability
in the ideal case, as $\hat{b}_k={\rm sgn}(m_k^t)$ at each update, 
where ${\rm sgn}(u)=u/|u|$ 
for $u \ne 0$. 
The computational cost for carrying out
this algorithm scales as $O(NK)$ per update when 
the one-dimensional integrations for assessing 
$m_k$ and $M_k$ are approximately evaluated 
with a fixed amount of time.

A tree approximation, which ignores time correlations of 
variables between successive updates, implies
that the macroscopic behavior of this algorithm is 
described by the natural iteration of the saddle point
equation of the 1RSB analysis. On the other hand, numerical 
experiments indicate that the discrepancy between the
trajectories of macroscopic variables 
and those of the 1RSB saddle point equation
is relatively large when the AT stability of the RS
solution is lost, 
in particular, during the transient stage.
However, the macroscopic property of the fixed point 
excellently accords with that predicted by 
the 1RSB solution. 

Support by Grants-in-Aid (Nos.~14084206 and~17340116) 
from MEXT/JSPS, Japan, is acknowledged. 
Useful discussion with A. Hatabu is appreciated.


\end{document}